\documentclass{aa}
\usepackage{epsfig}
\begin{document}
\def\h50{h$_{50}^{-1}${}}
\def\kms{km~s$^{-1}${}}
\thesaurus{11.01.2; 11.06.2; 11.09.2; 11.16.1; 11.17.3; 13.09.1}

\title{Near-infrared photometry of isolated spirals with and without an AGN.
II: Photometric properties of the host galaxies.
\thanks{Based on data obtained at: the European Southern
Observatory, La Silla, Chile, the T\'elescope Bernard Lyot, Calar Alto
Observatory, Las Campanas Observatory. Also based on observations made with 
the NASA/ESA Hubble Space Telescope, obtained from the data archive at the 
Space Telescope Institute}}

\author{
  I.~M\'arquez \inst{1}
\and
  F.~Durret \inst{2,3}
\and
  J.~Masegosa \inst{1}
\and
  M.~Moles \inst{4}
\and
  R.M.~Gonz\'alez Delgado  \inst{1}
\and
  I.~Marrero \inst{1}
\and
  J.~Maza \inst{5}
\and
  E. P\'erez \inst{1}
\and
  M. Roth \inst{6}
}
\offprints{I. M\'arquez (\sl{isabel@iaa.es}) }
\institute{
	Instituto de Astrof\'\i sica de Andaluc\'\i a (C.S.I.C.), 
Apartado 3004 , E-18080 Granada, Spain
\and
	Institut d'Astrophysique de Paris, CNRS, 98bis Bd Arago, 
F-75014 Paris, France 
\and 
	DAEC, Observatoire de Paris, Universit\'e Paris VII, CNRS (UA 173), 
F-92195 Meudon Cedex, France 
\and 
	Instituto de Matem\'aticas y F\'\i sica Fundamental (CSIC),
  Madrid, Spain and Observatorio Astron\'omico Nacional, Madrid, Spain
\and 
	Departamento de Astronom\'\i a, Universidad de Chile, Casilla 36D, 
Santiago, Chile
\and
    Observatories of the Carnegie Institution of Washington, 813 Barbara 
Street, Pasadena, CA91101
}

\date{Received,  ; accepted,}

\authorrunning{I. M\'arquez et al.}  

\titlerunning{Photometric properties of host galaxies of isolated
spirals with and without an AGN.}

\maketitle

\begin{abstract}

We present here the analysis of morphological and photometric
properties of a sample of isolated spirals with (18) and without (11)
an active nucleus, based on near-infrared imaging in the J and K'
bands (Paper I).  The aim of that comparative analysis is to find the
differential properties that could be directly connected with the
phenomenon of nuclear activity. We stress the importance of using
isolated objects for that purpose. Our study shows that both sets of
galaxies are similar in their global properties: they define the same
Kormendy relation, their disk components share the same properties,
the bulge and disk scale lengths are correlated in a similar way, bar
strengths and lengths are similar for primary bars. Our results
therefore indicate that hosts of isolated Seyfert galaxies have bulge
and disk properties comparable to those of isolated non active
spirals. Central colors (the innermost 200 pc) of active galaxies are
redder than the centers of non active spirals, most probably due to
AGN light being re-emitted by the hot dust and/or due to circumnuclear
star formation, through the contribution of giants/supergiants.

Central to our analysis is the study of the possible connection
between bars and similar non axisymmetric structures with the nuclear
fuelling. We notice that only one of the Seyfert galaxies in our
sample, namely ESO 139-12, does not present a primary bar. But bars
are equally present in active and control objects. The same applies to
secondary bars. Not all the active galaxies we have observed have
them, and some control galaxies also present such central
structures. Secondary central elongations (associated with secondary
bars, lenses, rings or disks) may be somewhat different, but this
result should be confirmed with larger samples.  We note that
numerical models indicate that such secondary bars are not strictly
necessary to feed the central engine when a primary bar is
present. Our results show that down to scales of 100-300 pc, there are
no evident differences between active and non active spiral galaxies.

\end{abstract}

\keywords{galaxies: active - galaxies: fundamental parameters -
galaxies: photometry -
infrared: galaxies }

\section{Introduction}

Many studies have been devoted to analyze the properties of Seyfert
host galaxies in order to understand the fuelling processes taking
place in active galactic nuclei (AGN).  At large scale, the fuelling
of the active nucleus is supposed to be due to the transport of gas
to the central region; this mechanism seems to be
connected to the presence of a bar, which provides the
non axisymmetric potential invoked in theoretical works (Simkin et
al. 1980, Shlosman et al. 1989, Barnes \& Hernquist 1991).  One of the
mechanisms proposed to drive the gas to the very central regions of
barred galaxies is that of nested bars (Shlosman et al. 1989; Friedli
\& Martinet 1993; Combes 1994; Heller \& Shlosman 1994), which has
been recently proved to fuel molecular gas into an intense central
starburst in NGC 2782 (Jogee et al. 1999) and in a Seyfert 2 galaxy,
Circinus (Maiolino et al. 1999).

Nevertheless, from the observational side, recent studies (Moles et
al. 1995; Ho et al. 1997; Hunt et al. 1999a) based on optical data
conclude that barred galaxies are equally found among active and
non active galaxies. Therefore, large scale bars are not the specific
property that identifies the family of active spiral galaxies.  NIR
imaging is more reliable in determining the overall mass distribution
in galaxies and it has been shown to be more efficient to detect bars
(Mulchaey et al. 1997; Seygar \& James 1998).  From their NIR imaging
analysis, McLeod \& Rieke (1995) and Mulchaey \& Regan (1997) find no
evidence for a significant excess of bars in Seyfert galaxies. But a
debate still exists on this matter: the analysis by Knapen et
al. (2000) of high resolution NIR images by Peletier et al. (1999)
points to an excess of bars among Seyferts at a 2.5$\sigma$ level,
attributing their different result to their better spatial resolution
and better matching of active and control samples than in previous
works.

A different approach to the problem, also adopted here, consists in
deriving detailed information on a number of selected objects, instead
of performing statistics on large samples.  Regan \& Mulchaey (1999)
analyze the HST high resolution dust morphology of 12 Seyferts,
searching for central bars. The non-ubiquity of such bars led them to
conclude that strongly barred potentials cannot be the only mechanism
for driving gas into the nucleus; they propose central spiral
dust lanes as an alternative method.  In the same vein, Martini \&
Pogge (1999) analyze HST images of 24 Seyfert 2s and find nuclear
spirals in 20 of them but only 5 with nuclear bars, concluding that
nuclear spirals may be the channel to feed gas into the central
engines.

Following this approach, we have chosen to analyze what the
similarities and/or differences are between active and non active
spirals.  Since gravitational interaction has been invoked to be very
efficient to induce the formation of bars or any other non
axisymmetric component, we take the approach of only selecting
isolated objects, in order to avoid the bias introduced by not taking
into account the environmental characteristics of the considered
galaxies.  The DEGAS project (Dynamics and Nuclear Engine of Galaxies
of Spiral type) aims at extending the analysis by Moles et al.  (1995)
and addresses the morphology of the galaxies, including optical and IR
images, and the kinematics of the stars and gas, through long slit,
high resolution spectroscopy.  In this paper we present the analysis
of the NIR imaging, which is particularly important because it allows:
1)~to separate the various components (the bulge, disk, bar(s) and
spiral arms) with the smallest contribution of the active nucleus;
2)~to detect and characterize the properties of bars and other
structures close to the nucleus, such as bars within bars, elongated
disks, rings or lenses, traced by the old stellar population.  We
present the analysis based on the data set presented in M\'arquez et
al. (1999), and compare the properties of active and non active
galaxies in our sample, and those of our sample galaxies with those
found in similar studies based on other samples.

Our sample is briefly described in Section \ref{sample}. The results
on the photometric decomposition and the bar properties are discussed
in Sections \ref{decomposition} and \ref{bars}. The discussion and
conclusions are given in Section \ref{discussion}.

\section{The sample and data}\label{sample}

The infrared imaging data for 18 active objects and 11 non active
galaxies are described in M\'arquez et al. (1999, hereafter Paper I).
The active galaxies have been chosen with the following criteria:
(a)~Seyfert 1 or 2 from the V\'eron-Cetty \& V\'eron (1993) catalogue;
(b)~with morphological information in the RC3 Catalogue; (c)~isolated,
in the sense of not having a companion within 0.4 Mpc (H$_0$=75
km/s/Mpc) and cz$<$500 km/s, or companions catalogued by Nilson
without known redshift; (d)~nearby, cz$<$6000 km/s; and
(e)~intermediate inclination (30 to 65$^\circ$). The non active  sample
galaxies have been selected among spirals verifying the same
conditions (b), (c), (d) and (e), and with morphologies (given by the
complete de Vaucouleurs coding, not just the Hubble type) similar to
those of the active spirals. Thus, all the galaxies in our sample are
isolated, in the sense of avoiding possible effects of interactions
with luminous nearby galaxies.

In Paper I we already stressed that these non active galaxies are well
suited to be used as a control sample. For each object we give in
Paper I: the image in the K' band, the sharp-divided image (obtained
by dividing the observed image by a filtered one), the difference
image (obtained by subtracting a model to the observed one), the J-K'
color image, the ellipticity and position angle profiles, the surface
brightness profiles in J and K' and their fits by bulge+disk models
and the J-K' color gradient.

The mean resolution of our images is about 1 arcsecond, corresponding
to a physical resolution between 100 and 300 parsecs for the closest
and the more distant galaxy respectively. This resolution implies that
we are able to map the region where the dynamical resonances are
expected to occur (see for instance P\'erez et al. 2000) and is
therefore well suited for our purposes.

\section{Results of the photometric bulge+disk 
decomposition}\label{decomposition}

\begin{table}[h]
\caption[ ]{Average bulge and disk effective surface brightnesses and
radii in J and K', and J-K' colors in the various components, for
active and non active galaxies. The corresponding 1$\sigma$
dispersions are also given.}
\begin{tabular} {lrr}
\\
\hline
&Active & Control\\
\\
\hline
$\mu_{bulge}^{K'}$   & 15.24 $\pm$ 1.87 & 15.73 $\pm$ 2.36\\ 
log R$_{bulge}^{K'}$     & -0.26 $\pm$ 0.42 & -0.24 $\pm$ 0.61\\
$\mu_{bulge}^{J}$    & 16.97 $\pm$ 0.49 & 16.74 $\pm$ 1.83 \\
log R$_{bulge}^{J}$      & -0.10 $\pm$ 0.44 & -0.25 $\pm$ 0.53\\
$\mu_{disk}^{K'}$    & 18.57 $\pm$ 0.62 & 18.94 $\pm$ 0.58\\
log R$_{disk}^{K'}$      &  0.68 $\pm$ 0.17 &  0.67 $\pm$ 0.18\\ 
$\mu_{disk}^{J}$     & 19.54 $\pm$ 0.29 & 20.30 $\pm$ 0.88\\
log R$_{disk}^{J}$       &  0.74 $\pm$ 0.23 &  0.60 $\pm$ 0.28\\
$a_1$&4.27 $\pm$ 1.59 &  4.94 $\pm$ 1.97 \\
$\epsilon_1$&0.35 $\pm$ 0.17 & 0.45 $\pm$ 0.18 \\
(J-K')$_{bulge}$     &  1.02 $\pm$ 0.70 &  1.25 $\pm$ 0.50\\
(J-K')$_{disk}$      &  0.78 $\pm$ 0.36 &  0.92 $\pm$ 0.33\\
(J-K')$_{prim. bar}$ &  1.03 $\pm$ 0.07 &  1.09 $\pm$ 0.07\\
(J-K')$_{centre}$    &  1.65 $\pm$ 0.44 &  1.14 $\pm$ 0.22\\ 
\hline
$a_2$& 1.28 $\pm$ 0.51 &  0.97 $\pm$ 0.23 \\
$\epsilon_2$&0.22 $\pm$ 0.16 & 0.28 $\pm$ 0.08 \\
(J-K')$_{prim. bar}$ &  1.02 $\pm$ 0.03 &  1.12 $\pm$ 0.06\\
(J-K')$_{sec. bar}$  &  1.14 $\pm$ 0.10 &  1.29 $\pm$ 0.19\\
(J-K')$_{centre}$    &  1.50 $\pm$ 0.35 &  1.09 $\pm$ 0.25\\
\hline
\\
\protect\label{decomp}
\end{tabular}
\begin{scriptsize}

Note: Surface brightnesses ($\mu$) are given in mag/$\sq\arcsec$ and
sizes in kpc. $a_i$ and $\epsilon_i$ are deprojected values for the
bar ($i$=1 for the primary and $i$=2 for the secondary) semi-major
axis and ellipticity ($\epsilon$ = $1-b/a$). The second part of the table
refers to galaxies with both primary and secondary bars.
\end{scriptsize}
\end{table}

The surface brightness profiles derived for all the galaxies together
with the best bulge/disk decomposition (1D exponential disk and de
Vaucouleurs bulge) are displayed in Figs. 1-28 f-g, and the
corresponding observed parameters are given in Table 3 of Paper~I.
Note that several galaxies (namely the active galaxies NGC 3660 and
NGC 5728, and the non active galaxies NGC 151, NGC 2811 and NGC 3571)
extend notably further out than the infrared images, so the resulting
bulge+disk decomposition cannot be satisfactory.  Full 2D fits have
not been considered since, as already noticed by de Jong (1996b), they
would not be better than 1D fits for the purposes of the present
study.

The average bulge and disk effective surface brightnesses and
effective radii in J and K' as well as J-K' colors are given in Table
\ref{decomp}. Magnitudes have been corrected for galactic extinction,
inclination (assuming transparent disks) and redshift (we have applied
the K correction following Hunt et al. 1997).  We note that, at face
values, some differences are found between active and control
galaxies. However, the application of non-parametric tests shows that
those differences are not significant.

With respect to the bulge component, we recall that the parameters
derived from the photometric decomposition are more critically
dependent on the fitting procedure (see de Jong 1996a; Moriondo et
al. 1996); in addition to this, the contribution of the AGN in active
galaxies is not easy to extract, so bulge parameters can be less
accurate. With these caveats in mind, the differences in the average
values reported in Table \ref{decomp} are not conclusive; the
Kolmogorov-Smirnoff (KS) test gives a probability greater than 98\%
for both samples to have the same magnitude, surface brightness and
scale-length.

\begin{figure}[ht]
\centerline{\psfig{figure=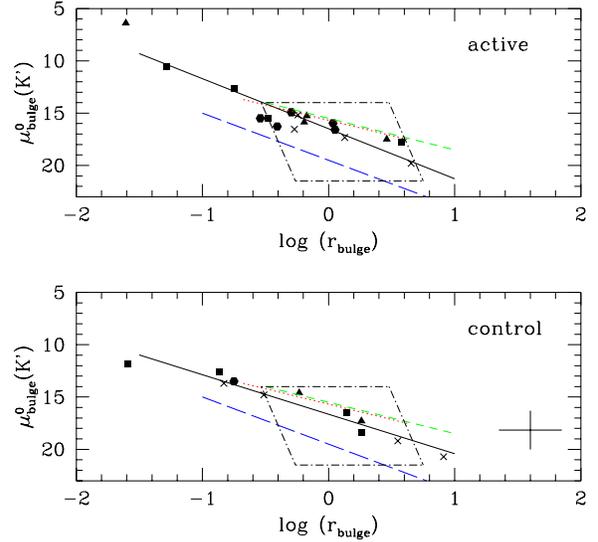,height=8cm}}
\caption[ ]{K' band bulge equivalent surface brightness (in mag/$\sq\arcsec$)
vs. equivalent radius (in kpc) for active (top) and control (bottom) 
galaxies. The different
symbols are: triangles for Sa, squares for Sab, circles for Sb and
crosses for Sbc. The four lines correspond to the Kormendy relations
found for our data (full line), by Moriondo et al. (1998) (dotted line) 
and by Hunt et al. (1999a) (short dashed line); the Xanthopoulos (1996) 
relation in the I band is given for comparison (long dashed line). 
The box indicates the region occupied by spiral galaxies from de Jong 
(1996b). Typical errors are represented by the cross in the bottom pannel.}  
\protect\label{bulboac}
\end{figure}

In Fig. \ref{bulboac} the bulge equivalent surface brightness is
plotted as a function of equivalent radius (i.e. the Kormendy 1977
relation for bulges) for active and control galaxies. Different
symbols were used for the various morphological types. However, since
the number of galaxies in our sample is small, it is not possible to
define Kormendy relations for the various morphological types.  The
best fits are $\mu _{bulge}=(16.64 \pm 0.24) + (3.77\pm 0.33)\ log\
r_{bulge}$ and $\mu _{bulge}=(16.48 \pm 0.30) + (4.79\pm 0.48)\ log\
r_{bulge}$ for control and active galaxies respectively ($\mu
_{bulge}=(16.52 \pm 0.21) + (4.29\pm 0.32)\ log\ r_{bulge}$ when the
two samples are taken together).  The Kormendy relations obtained for
similar data in the near infrared by other authors are also drawn in
these figures.  
Our data are in
good agreement with these relations, and extend to somewhat smaller
bulges.

The Xanthopoulos (1996) relation in the I band for Seyfert 1
vs. Seyfert 2 galaxies is also given for comparison (27
galaxies). Although it is displaced vertically relative to the
relations found in K', as expected, it shows a comparable
slope. Therefore, considering that bulge scale-lengths should 
essentially be the same from I to K' for a given galaxy (Evans 1994; see
also Hunt et al. 1999a and below), this implies that the bulge
component of Seyfert spirals would have an essentially uniform I-K'
color ((I-K') $\approx$ 2.7).

The region occupied by the bulges obtained by de Jong (1996b) is also
plotted in Fig. \ref{bulboac}.  The bulge parameters for our sample
galaxies define a narrower relation than that found by de Jong,
probably because we select only the most isolated objects. The same
conclusion was found by M\'arquez \& Moles (1999) when comparing the
bulge parameters of their isolated spirals (M\'arquez \& Moles 1996)
with those of de Jong in the B band and Baggett et al. (1998) in the V
band.

The main result is therefore that no difference is found between the
bulges of active and control galaxies, in agreement with Hunt et
al. (1999a).  It appears that the bulges of active and control galaxies
have totally similar structural properties.

\begin{figure}
\centerline{\psfig{figure=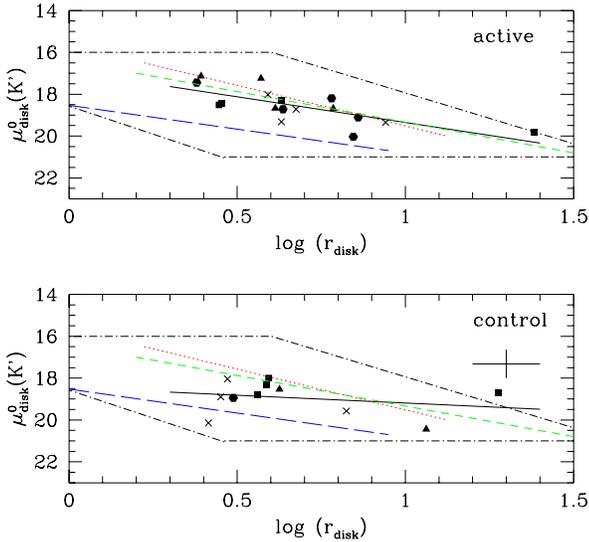,height=8cm}}
\caption[ ]{Same as Figure \ref{bulboac} but for the disk component.}
\protect\label{discoac}
\end{figure}

With respect to the disk component, the disk surface brightness as a
function of its equivalent radius is plotted in Fig. \ref{discoac} for
active and control galaxies. The best fits are $\mu _{disk}=(16.89 \pm
0.43) + (2.46\pm 0.60)\ log\ r_{disk}$ and $\mu _{disk}=(18.44 \pm
0.67) + (0.75\pm 0.94)\ log\ r_{disk}$ for active and control galaxies
respectively.  The slope of the relation is heavily dependent on the
very large and uncertain value found for IC~454. 
If that point is taken out of the relation, we find for control
galaxies $\mu _{disk}=(16.82 \pm 0.57) + (3.21\pm 0.86)\ log\
r_{disk}$. Therefore, both control and active galaxies seem to share
the same relation ($\mu _{disk}=(16.98 \pm 0.35) + (2.54\pm 0.50)\
log\ r_{disk}$ for both samples simultaneously).

As for bulges, our data on disks are in good agreement with the
relations found by Moriondo et al. (1998) and Hunt et al. (1999a) for
normal galaxies. The relation is again narrower for our isolated
galaxies than for de Jong's sample, as reported in M\'arquez \& Moles
(1999). The relation found by Xanthopoulos (1995) in the I band is
parallel to ours; considering the same scale length for the different
bands (as for the bulge, also see below), this would imply an
essentially uniform value for the disk color of isolated Seyfert
spirals ((I-K') $\approx$ 1.5).

Average values for the J-K' colors of bulges and disks are given in
Table \ref{decomp}.  Bulge colors agree with those by Moriondo et
al. (1998) who find a mean (J-K)=1.06 $\pm$ 0.3 for a sample of 14
early-type non active spirals.  They also agree with the results by
Hunt el al. (1999a), who found (J-K)=1.04 and (J-K)=1.07 for the
bulges of Seyfert 1 and normal Sa galaxies, respectively.  Disk colors
occupy a narrower range, in good agreement with previous (J-K)
determinations (Hunt et al. 1997; Hunt et al. 1999a). There appears to
be a trend, with a large spread of values, for the bulges and disks of
active galaxies to have smaller J-K' values than the control sample,
but an application of the KS test results in a $>$ 99\% probability for
both samples to have the same disk color distribution.

Bulge equivalent radii in the J and K' bands follow the same relation
for both active and control galaxies: $r_{bulge}(K')=(0.915\pm
0.029)r_{bulge}(J)$.  With respect to the disk equivalent radii in the
J and K' bands: $r_{disk}(K')=(0.988\pm 0.026)r_{disk}(J)$.  Both
bulge and disk scale lengths appear to be essentially the same
in J and K'.

A comparison of the bulge and disk equivalent radii in the J and K'
bands indicates that, again,
both active and control galaxies show the same trend:
$r_{bulge}(K')=(0.12\pm 0.07)r_{disk}(K')$ and $r_{bulge}(J)=(0.17\pm
0.11)r_{disk}(J)$. This is in agreement with previous results by de
Jong (1996b) (mean $r_{bulge}/r_{disk}$ = 0.14 in K'), Courteau et
al. (1996) (mean $r_{bulge}/r_{disk}$ = 0.13 in R), and Graham \&
Prieto (1999) (mean $r_{bulge}/r_{disk}$ = 0.24 in K' for early
spirals). A weak correlation between $r_{bulge}$ and $r_{disk}$ is
reported by Seygar \& James (1998).

\begin{figure}[h]
\centerline{\psfig{figure=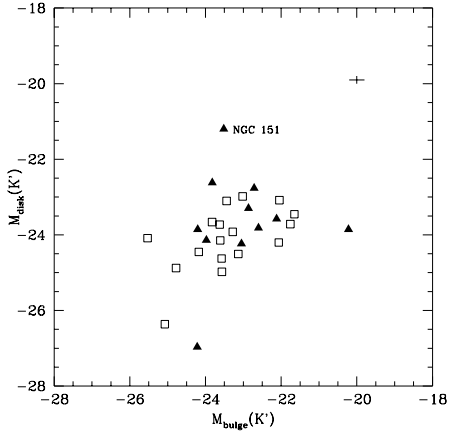,height=6cm}}
\caption[ ]{Relation between bulge and disk absolute magnitudes in K'.
Active and non active galaxies are indicated with empty squares and full 
triangles respectively. Typical errors are represented by the cross.}
\protect\label{MbMd}
\end{figure}

In Fig. \ref{MbMd} we show the relation between the bulge and disk
absolute magnitudes in K'. Our results are in good agreement with
those by de Jong (1996b), considering that we deal with early-type
spirals.

Within the spread of values we find no apparent differences in the
photometric parameters between the host galaxies of the active and
control samples. This is in good agreement with previous results,
except for the disk surface brightnesses of Seyfert galaxies. Bulges
of Seyferts in our sample cover the same region as those by Bender et
al. (1992) and Andreadakis et al. (1995).  The disks of both control
and active galaxies are similar to those of normal galaxies in Hunt et
al. (1999a; taken from de Jong 1996b and Moriondo et al. 1998) as seen
in Fig. \ref{discoac}, and also agree with those by Andreadakis et
al. (converted to the K band as in Hunt et al. 1999a). Nevertheless,
the disk surface brightness of Seyfert galaxies selected from the CfA
sample by Hunt et al. (1999a) and from the 12$\mu$m sample by Hunt et
al. (1999b) are about 1 magnitude brighter in K than those of normal
early type spirals.  However, it has to be noted that their control
samples were not specially designed to match the active sample; in
particular, some clearly interacting systems are found among the
active sample, so interactions may have an effect on the enhanced
surface brightness they obtain.

This raises the question of how well the control sample used for
comparison matches the active sample. Very recently, Knapen et
al. (2000) have reported a higher bar percentage for active spirals with
respect to non active ones.
We stress that, unlike other authors, we are avoiding the possible
effects of any strong interaction on the disk surface brightness. In
particular, if our isolation criteria are applied to select only the
isolated objects among those analyzed by Knapen et al., their
conclusions change: the (small) resulting samples of active and
control galaxies (13 and 11 with morphological information on the
presence or absence of a bar) do show the same percentage of barred
galaxies (8 and 7 galaxies, respectively). This result reinforces the
importance of not including objects that could be suffering from a
gravitational interaction (as it is the case, for instance, for NGC
7469, which is included in Knapen's sample and is known to reside in
an isolated pair; see, for instance, M\'arquez \& Moles 1994).  Since
the sample of active galaxies by Hunt et al. (1999a) also comes from
the CfA sample, the same considerations should apply and could explain
the higher disk surface brightness they found as a result of including
interacting objects. This point should be statistically confirmed by
using a larger sample of active and non active {\bf isolated} spirals.

\section{Bars and central properties}\label{bars}

We discuss here the properties of primary and secondary bars in the
studied galaxies. We remind the reader that we refer to secondary bars
every time a central elongation is detected, be it a bar, a lens, an
inclined disk or a ring. Whether that elongation is actually a bar is
a question that will be addressed using kinematic data.

Observed bar lengths and ellipticities are given in Paper I. They
have been obtained from the ellipse fitting, corresponding to peaks in
ellipticity for constant PA.  All the discussion refers to deprojected
bar parameters (see Jungwiert et al. 1997).

\begin{figure}[h]
\centerline{\psfig{figure=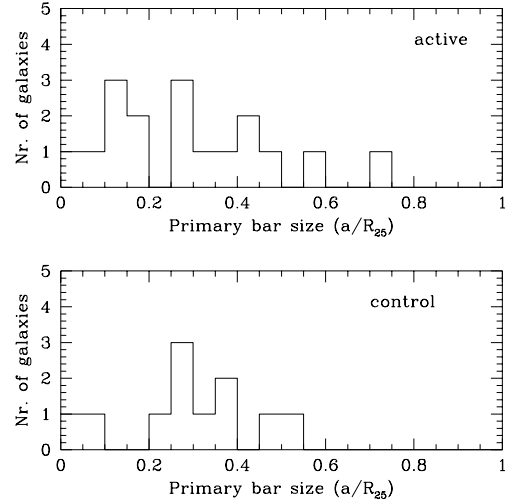,height=7cm}}
\caption[ ]{Distribution of relative sizes of the primary bars 
in active  (N=17) and control (N=11) galaxies.}  
\protect\label{histobar1}
\end{figure}

\begin{figure}
\centerline{\psfig{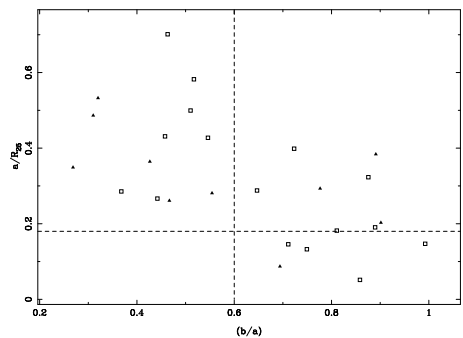}}
\caption[ ]{Relation between primary bar strength and length for
active (squares) and control (triangles) galaxies. The limits for
strong and long bars as defined by Martinet \& Friedli (1997) (see
text) are given as dashed lines.}  \protect\label{comp_chapelon}
\end{figure}

Regarding the primary, large scale bars, we find that the distribution
of sizes is similar for both active and control galaxies. The
same result is found when considering relative bar sizes (with respect
to the disk size), as shown in Fig. 5. Both histograms are rather
flat, and a KS test indicates that they are similar with a 
probability higher than 99.5\%.  The strength of the bars has been
parametrized with (b/a), with smaller values of (b/a) for
stronger bars as discussed by Martinet \& Friedli (1997), and Chapelon
et al. (1999). It has to be noted that their the criteria to
estimate bar lengths and ellipticities are not exactly the same as
ours. Nevertheless, we will be comparing active and control galaxies
for which the estimations come from the same procedure (ellipse
fitting, see above).  In Fig. 5 we have plotted the bar strength as a
function of the normalized length ($a/R_{25}$). Our results agree with
those presented by the quoted authors. The comparison for active and
control galaxies, with mean strengths of $(b/a)$ = 0.68 $\pm$ 0.19 and
0.51 $\pm$ 0.24, and mean lengths of $(a/R_{25})$ = 0.29 $\pm$ 0.18
and 0.32 $\pm$ 0.13 respectively, don't show any significant
difference. Martinet \& Friedli (1997) defined strong bars for $(b/a)$
$\leq$ 0.6 and long bars for $(a/R_{25}) \geq 0.18$. According to this
definition, Fig. \ref{comp_chapelon} also shows that the whole range
from weak to strong bars is equally represented for both active and
control galaxies. The majority of barred galaxies in both samples
harbor long bars (this seems to be specially the case for control
galaxies).

The fraction of detections of secondary bars is similar for both active and
control galaxies. As suggested by Friedli \& Martinet (1993), the
formation of the secondary bar is mainly driven by the main bar, so in
this case we calculate relative sizes of secondary bars with respect
to primary ones. The distribution of the relative sizes is shown in Fig.  7. 
For active galaxies the distribution is more extended,
but the differences are not significant. We stress the fact that the
presence of secondary bars is not exceptional in non active galaxies.
									 
Finally, the consideration of the primary and secondary bars together
doesn't show any correlation between the properties of both
structures, or with the effective radii of the bulges and disks, in
agreement with previous findings (Seygar \& James 1998). The only hint
we find is for a positive correlation between the sizes of the primary
and secondary bars, but the scatter is too large and the sample
small. In any case this would not be a difference between active and
non active galaxies, since it would be present for the whole sample
here.

\begin{figure}
\centerline{\psfig{figure=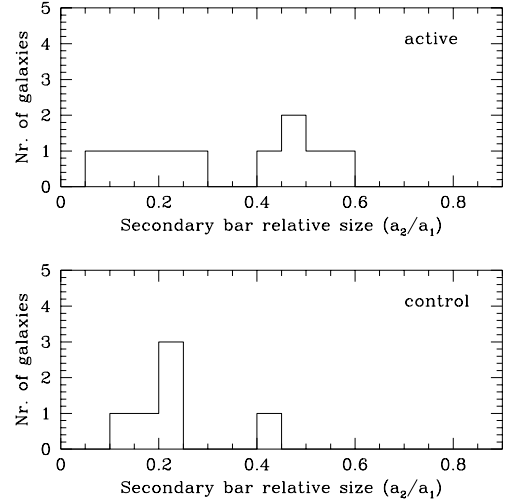,height=7cm}}
\caption[ ]{Distribution of relative (with respect to the primary bar) 
sizes of the secondary bars in active  (N=10) and control (N=6) galaxies.}  
\protect\label{histobar2}
\end{figure}

\begin{figure}
\centerline{\psfig{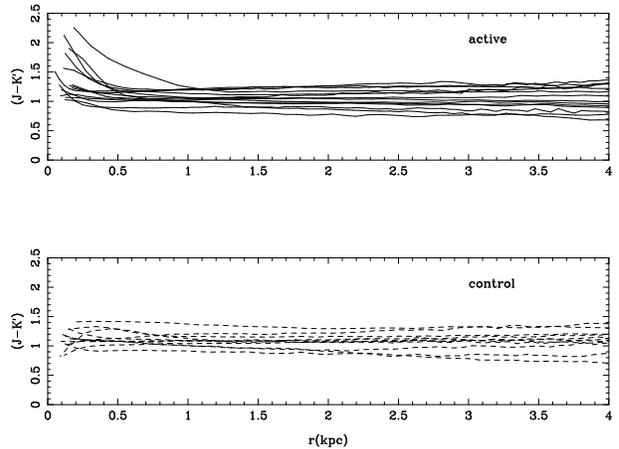}}
\caption[ ]{(J-K') color gradients for active (top) and control (bottom) 
isolated spirals.}  
\protect\label{gradcol}
\end{figure}

\begin{figure}
\centerline{\psfig{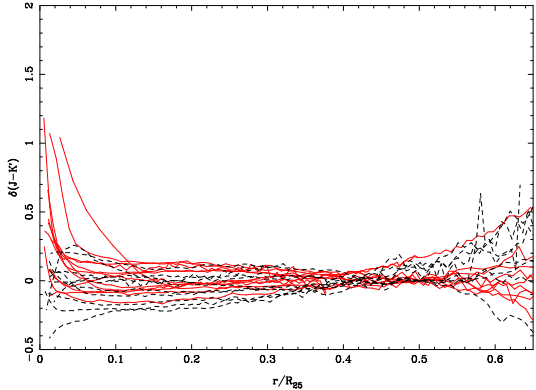}}
\caption[ ]{Normalized color gradients 
($\delta$(J-K')(u) = (J-K')(u) - (J-K')$(0.5)$, with $u$=r/R$_{25}$) 
for active (full line) 
and control (dashed line) galaxies.}  
\protect\label{gradcol_spl}
\end{figure}

Color gradients as a function of distance to the center (in kpc) are
shown in Fig. \ref{gradcol}.  Disk colors are similar for both
samples. Central colors of control galaxies agree with those of
typical spirals determined by Griersmith et al. (1982) and Forbes et
al. (1992).  We have computed a normalized color gradient
$\delta$(J-K')(u) = (J-K')(u) - (J-K')$(0.5)$, where $u$ is the
distance to the center in units of R$_{25}$, $u$=r/R$_{25}$ so that
(J-K')$(0.5)$ is the color at a fixed relative distance of
0.5$\times$R$_{25}$, which corresponds to the region were color
gradients are reliable (see Paper I). This color gradient is plotted
in Fig. \ref{gradcol_spl} for active and control galaxies.  Note than
only three galaxies harbor bars longer than 0.5$\times$R$_{25}$
(namely NGC 4785, NGC 5728 and NGC 151), which have been excluded from
the plot, so Fig.  \ref{gradcol_spl} shows colors inside the bar
region. It can be seen that $\delta$(J-K')(0.5) = 0 within the
photometrical errors. $\delta$(J-K')(u) values depart from
$\delta$(J-K')(u)=0 further out, mostly due to the background
subtraction difficulties reported in Paper I. The differences in the
behaviours of central regions of active and control galaxies are well
visualized in this plot.  In the innermost 0.1$\times$R$_{25}$, active
galaxies are generally redder than control galaxies (the reddest
active galaxy is $\approx$1 magnitude redder than the reddest control
galaxy).  We also note that three control galaxies (namely NGC 2712,
NGC 3835 and NGC 6155) show bluer colors inside 1 kpc, whereas none of
the active galaxies becomes bluer in the center. Shaw et al. (1995)
found that 19 out of a non-complete sample of 32 large barred galaxies
show blue nuclei. Among them, only four were Seyferts.  The result of
non active galaxy with blue nucleus among isolated spirals should be
confirmed with larger samples.

The average J-K' colors of the central region (the innermost 200 pc)
and of the primary bar are given in Table \ref{decomp} for all the
galaxies of our sample that have a bar. The colors of the primary bars
appear to be similar in active and non active galaxies, and also to be
similar to the J-K' value of the central region of the control
sample. On the other hand, the central regions of the active sample
have larger J-K' values than the control one, in agreement with
Kotilainen \& Ward (1994). This difference cannot be explained as due
to differential contributions of the nebular continuum in the J and K'
bands produced by the NLR of active galaxies. \footnote{We have
estimated that the central J-K' colors change only by less than 10\%~
if the color of each active galaxy is corrected for the nebular
continuum contribution, estimated from the H$\alpha$ or $H\beta$
fluxes of the nucleus reported in the literature.} It could be
understood as coming either from an important contribution of active
star forming regions usually found associated with the presence of the
active nucleus in active galaxies (Rodr\'{\i}guez-Espinosa et
al. 1987; Wilson 1988; Hunt \& Giovanardi 1992) through the
contribution of giants/supergiants (Shaw et al. 1995) or from dust
re-emission in a dusty starburst or from the AGN illumination (Seygar
\& James 1999).

\section{Discussion and Conclusions}\label{discussion}

Our analysis is intended to detect the differential properties between
isolated active and non active spiral galaxies. We have considered
here the large scale, global properties as well as the detailed
morphology of the central regions, down to 100-300 pc. Our results
show that there are no sizeable differences between active and non
active isolated spiral galaxies in the volume limited sample
considered here. By this, we mean that the global properties are
similar, and none of the detected structures is exceptionally present
or absent in one of the groups.  Our results concerning the global
properties, refer to a rather small sample and are in general
agreement with previous ones, reinforcing our conclusion that hosts of
isolated Seyfert galaxies have bulge and disk properties comparable to
those of isolated non active spirals. In particular:

- both samples define the same (Kormendy) relation between $\mu_{eff}$
and $r$ for bulges ;

- disk components also share the same properties. This contradicts the
result by Hunt et al. (1999a, 1999b) that Seyfert disks are about 1
magnitude brighter than the disks of non active spirals. This
discrepancy may be explained in terms of a possible contamination by
interacting objects in their sample;

- bulge and disk scale lengths are correlated, with $r_{bulge} \approx
0.2\ r_{disk}$;

- central colors of active galaxies are redder than the centers of
non active spirals, most probably due to the AGN light re-emitted by
the hot dust and/or to the presence of active star formation in
circumnuclear regions.

It is generally admitted that the mechanism responsible for the
transport towards the center and the possible onset of nuclear
activity could be related with the presence of bars or other non
axysimmetric structures. Regarding the primary bars, we know that the
fraction of barred galaxies is not different among active and non
active spiral galaxies (Mc Leod \& Rieke 1995, Moles et al. 1995, Ho et
al. 1997, Mulchaey \& Regan 1997, Hunt \& Malkan 1999).  We notice that
only one of the active galaxies in our sample, namely ESO~139-12, does
not harbor a primary bar. This exception could however be of some
interest since it raises the question of what kind of mechanism could
produce non circular gas motions in an isolated object with no
large-scale bar.  Our results also show that primary bars have the
same mean strength and length in both families.

The difference could reside in the properties of secondary central
elongations (bars, inclined disks or rings, see Paper I). The complete
analysis has to include photometric and kinematic information, in
order to see what specific processes are taking place and whether they
differ between active and control galaxies. But even with only the
morphological information we report here, we can already derive some
conclusions, keeping in mind the limitations imposed, in particular by
the resolution reached by the IR images we have studied.  Our result
is that, down to scales of 100-300 pc, the detection rate of secondary
bars is not different for active and control galaxies. Admittedly our
sample is too small to allow strong statistical conclusions. But the
fact is that the presence of secondary bars is not exceptional among
non active galaxies. On the other hand, there is a number of active
spirals in our sample with no detected secondary bar.

There remains the question of whether differences would be found at
smaller scales, in particular those related to the presence of
different nuclear structures. Even if this point is out of the scope
of this paper, we note that HST images are only available for one of
the galaxies in the control sample, so no comparison between active
and non active galaxies can be attempted at this stage. In this
respect, we stress that the feedback from numerical simulations is
crucial to understand the mechanisms that are at the origin of the
nuclear activity, and the scale at which they should operate. Thus,
nuclear bars have been searched for in active galaxies because they
were needed to provide the second step to transport gas from the
circumnuclear region to the AGN (Shlosman et al. 1989; Friedli \&
Martinet 1993; Combes 1994; Heller \& Shlosman 1994).  As stated by
Maiolino et al. (2000), the interpretation of the data depends on both
the spatial resolution and the colours used to detect nuclear bars. In
addition to this, the interpretation of the data within numerical
model predictions is not straightforward. Regan \& Mulchaey (1999) and
Martini \& Pogge (1999) indeed searched for straight dust lanes,
considered as tracers of nuclear bars; however, the gas morphology
strongly depends on the bar pattern speed, with straight shocks
occurring only when the bar is rapidly rotating (Maciejewski \& Sparke
1999).  Therefore, slow rotating nuclear bars could also lead to the
``spiraling'' structure described by Regan \& Mulchaey (1999) and
Martini \& Pogge (1999), therefore increasing the number of Seyfert
galaxies hosting nuclear bars.

For Seyfert galaxies with no nuclear bar, di\-ffe\-rent mechanisms have to
be invoked, such as the presence of nuclear spirals (Martini \& Pogge
1999; Englmaier \& Shlosman 2000).  Recent numerical simulations are
beginning to predict an efficient enough transport of matter to the
center in spiral galaxies with only a primary bar (Maciejewski,
private communication), as should be the case for barred Seyferts with
no nuclear bar.

The detailed analysis of such nuclear features requires in any case,
in addition to the morphological information, a full kinematic
characterization. In particular, in the case of NGC~6951 (P\'erez et
al. 2000) we have shown that no nuclear bar is detected in the HST
images, but that there seems to be a nuclear spiral; this spiral
structure is most probably residing in a nuclear disk, kinematically
decoupled from the large-scale disk of the galaxy; the strong
molecular gas accumulation could have destroyed a pre-existing nuclear
bar and eventually may result in the dilution of the primary
bar.\footnote{We note that the question of gas concentration cannot be
addressed through molecular gas mapping since at this stage, no CO
maps are available for the remaining galaxies in our samples.} The
kinematic characterization of the central regions is expected to shed
light on the question of how the central kinematics are related to the
fuelling mechanisms in active galaxies.  In the same fashion as for
NGC~6951, we are now analyzing in detail the morphological and
kinematical properties of the galaxies in our sample.

\begin{acknowledgements}
I.~M\'arquez acknowledges financial support from the Spanish
Ministerio de Educaci\'on y Ciencia (EX94-8826734). This work is
financed by DGICyT grants PB93-0139, PB96-0921, PB98-0521 and PR95-329. 
Financial support to
develop the present investigation has been obtained through the Junta
de Andaluc\'{\i}a, the French-Spanish grants HF1996-0104 and
HF1998-0052, from the Picasso program of the French Ministry of
Foreign Affairs, and from the Chilean-Spanish bilateral agreement
CSIC-CONICYT 99CL0018. We also acknowledge financial support from
INSU-CNRS for several observing trips.
\end{acknowledgements}

\end{document}